\documentclass[aps,twocolumn,groupedaddress,superscriptaddress,showpacs]{revtex4}

\usepackage{graphicx}
\usepackage{colortbl}

\makeatletter
\def\btt#1{\texttt{\@backslashchar#1}}%
\DeclareRobustCommand\bblash{\btt{\@backslashchar}}%
\makeatother

\begin{document}


\title{Phase-change memory function of correlated electrons in organic conductors} 

\author{H.~Oike}
\affiliation{RIKEN Center for Emergent Matter Science (CEMS), Wako 351-0198, Japan}

\author{F.~Kagawa}
\email{fumitaka.kagawa@riken.jp}
\affiliation{RIKEN Center for Emergent Matter Science (CEMS), Wako 351-0198, Japan}
\affiliation{CREST, Japan Science and Technology Agency (JST), Tokyo 102-0076, Japan}

\author{N.~Ogawa}
\affiliation{RIKEN Center for Emergent Matter Science (CEMS), Wako 351-0198, Japan}

\author{A.~Ueda}
\affiliation{Institute for Solid State Physics, The University of Tokyo, Kashiwa, Chiba 277-8581, Japan}

\author{H.~Mori}
\affiliation{Institute for Solid State Physics, The University of Tokyo, Kashiwa, Chiba 277-8581, Japan}

\author{M.~Kawasaki}
\affiliation{RIKEN Center for Emergent Matter Science (CEMS), Wako 351-0198, Japan}
\affiliation{Department of Applied Physics, University of Tokyo, Tokyo 113-8656, Japan}

\author{Y.~Tokura}
\affiliation{RIKEN Center for Emergent Matter Science (CEMS), Wako 351-0198, Japan}
\affiliation{Department of Applied Physics, University of Tokyo, Tokyo 113-8656, Japan}

\date{\today}

\begin{abstract}
Phase-change memory (PCM), a promising candidate for next-generation non-volatile memories, exploits quenched glassy and thermodynamically stable crystalline states as reversibly switchable state variables. We demonstrate PCM functions emerging from a charge-configuration degree of freedom in strongly correlated electron systems. Non-volatile reversible switching between a high-resistivity charge-crystalline (or charge-ordered) state and a low-resistivity quenched state, charge glass, is achieved experimentally via heat pulses supplied by optical or electrical means in organic conductors $\theta$-(BEDT-TTF)$_2$$X$. Switching that is one order of magnitude faster is observed in another isostructural material that requires faster cooling to kinetically avoid charge crystallization, indicating that the material's critical cooling rate can be useful guidelines for pursuing a faster correlated-electron PCM function.

\end{abstract}


\maketitle

	When liquid crystallizes upon cooling, minute crystallites nucleate with a finite probability and subsequently grow in size. Because these processes proceed in a limited temperature range below the melting temperature $T_{\rm m}$ \cite{Ref1}, crystallization can be kinetically avoided when the temperature range is quickly passed through, resulting in a metastable glassy state \cite{Ref2, Ref3}. Such crystal/glass formation kinetics underlies the operation of chalcogenide phase-change memory (PCM) \cite{Ref4, Ref5, Ref6, Ref7, Ref8, Ref9, Ref10}, which is widely used for optical data storage and is currently under development for next-generation nonvolatile memory applications: in the crystal-to-glass transformation, the local temperature is increased above the $T_{\rm m}$ using optical or electrical pulses, and subsequent rapid cooling leads to the formation of the glassy state; to switch back, a relatively long pulse with a moderate intensity is applied to anneal the glassy-form material at a temperature above the glass transition temperature $T_{\rm g}$ ($< T_{\rm m}$), and the crystallization proceeds while the pulse is exerted on the material. Data processing is conducted on the basis of a large difference in resistivity or optical reflectivity between the two forms. 

An exotic class of PCM that does not rely on melting of the crystal of atoms or molecules may be designed if liquid-, crystal-, and glass-like electronic states are all incorporated in a single solid; however, the unique memory concept that harnesses a quenched non-equilibrium state of matter as well as a thermodynamically stable state has thus far been demonstrated exclusively in an aggregation of atoms. In this Rapid Communication, we demonstrate PCM functions emerging from a charge-configuration degree of freedom in organic conductors $\theta$-(BEDT-TTF)$_2$$X$ [where BEDT-TTF denotes bis(ethylenedithio)tetrathiafulvalene and $X$ denotes an anion] [Fig.~1(a)]. The non-volatile switching between a high-resistivity charge-crystalline state and a low-resistivity quenched charge glass is achieved via both optical and electrical means. Our results indicate that charge-ordering systems can be potential candidates for a new class of PCM.


\begin{figure*}
\includegraphics[width=17cm]{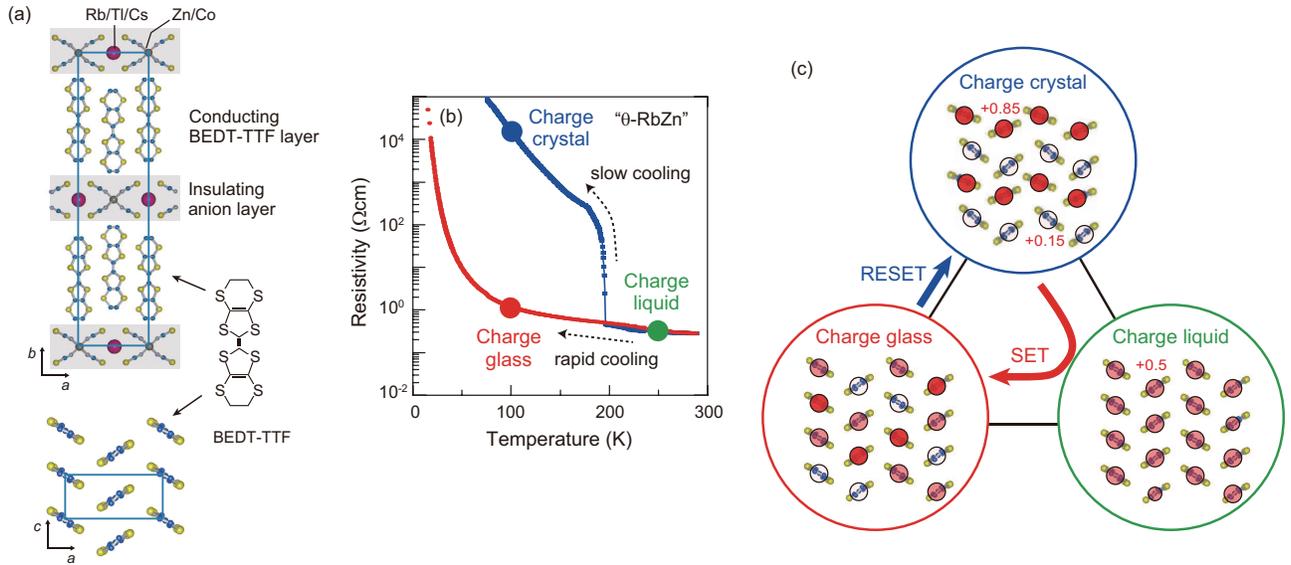}
\caption{\label{Fig1}
(Color online) (a) The crystal structure of $\theta$-(BEDT-TTF)$_2$$X$ [$X =$ RbZn(SCN)$_4$, TlCo(SCN)$_4$, and CsZn(SCN)$_4$]. The conducting BEDT-TTF layer viewed from the $b$-axis is also shown. The room-temperature space group is $I$222, and the unit cells are indicated by rectangles. 
(b) The temperature dependence of the resistivity during cooling with different temperature-sweeping rates in $\theta$-(BEDT-TTF)$_2$RbZn(SCN)$_4$ (denoted by $\theta$-RbZn). 
(c) A scheme for the rewritable switching of the correlated-electron phase-change memory implemented using $\theta$-(BEDT-TTF)$_2$$X$. The charge states at the temperatures indicated by closed dots in (b) are the charge liquid, the charge crystal, and the charge glass; their schematic structures are illustrated in (c). }

\end{figure*}

For demonstrating a PCM function emerging from correlated electrons, $\theta$-(BEDT-TTF)$_2$$X$ appears to be a paradigmatic system. $\theta$-(BEDT-TTF)$_2$$X$ is a layered material composed of the conducting BEDT-TTF layer and the insulating anion $X$ layer [Fig.~1(a)] \cite{Ref11}.  The transport properties for $X =$ RbZn(SCN)$_4$ (denoted as $\theta$-RbZn) and the schematics of the electronic states \cite{Ref12, Ref13, Ref14, Ref15, Ref16, Ref17} are summarized in Figs.~1(b) and 1(c), respectively; at high temperatures, the charges (holes) are itinerant in the BEDT-TTF layer, and the molecular valence is spatially uniform (+0.5). The charge state can therefore be referred to as liquid-like (``charge liquid''). At $\approx$195 K, the system undergoes a first-order transition into a Wigner-type charge-ordered state, in which hole-rich (+0.85) and hole-poor (+0.15) sites are arranged periodically (crystal-like). This charge ordering, or ``charge crystallization'', is accompanied by a structural change, and $\theta$-RbZn can therefore be viewed as a charge-lattice coupled system. When $\theta$-RbZn is cooled faster than the critical cooling rate $R_{\rm c}$ ($\approx 8$$\times$$10^{-2}$ K/s), the charge crystallization is kinetically avoided, and consequently, the charge liquid is frozen into a ``charge glass'' \cite{Ref16}, in which the molecular valence is broadly distributed from site to site \cite{Ref17}. Such cooling-rate-dependent bifurcation of the charge liquid is reminiscent of the situation in atomic/molecular liquids, including the chalcogenide PCM. Analogous to the chalcogenide PCM, therefore, we can envisage that by applying heat pulses with appropriate intensity and period, the high-resistivity charge crystal can be transformed into a low-resistivity charge glass through the charge liquid (``SET'' process), and moreover, the produced charge glass reverts to the charge crystal (``RESET'' process) [Fig.~1(c)].

\begin{figure}
\includegraphics[width=8.6cm]{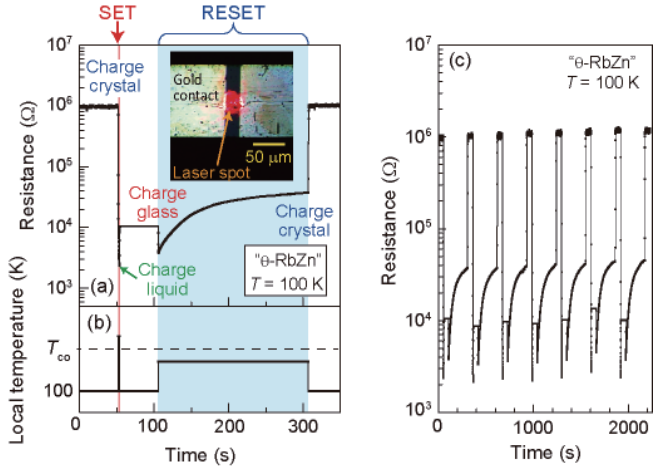}
\caption{\label{Fig2}
(Color online) (a, b) Single-cycle operation of the SET and RESET processes by optical means in $\theta$-RbZn mounted on a sample holder at 100 K. The two-probe resistance and the deduced local temperature at the laser spot are shown in (a) and (b), respectively, as a function of time. The hatches denote the duration of the SET and RESET laser irradiation. The optical microscope image of the sample setup is displayed in the inset to (a). 
(c) Repetitive operation of the correlated-electron phase-change memory function under the SET and RESET laser irradiation.}

\end{figure}

To verify the concept of the PCM function emerging from correlated electrons, we performed two-probe resistance measurements under the irradiation of a cw laser diode. Prior to this experiment, $\theta$-RbZn was slowly cooled to 100 K so that the entire sample could turn into a high-resistivity charge crystal. Figure~2(a) shows the typical phase-change behavior, along with the deduced local-temperature protocol at a laser spot [Fig.~2(b)]. In the SET process, a laser with an average power density of 110 W/cm$^{2}$ (1.4 mW on a 40-$\mu$m$\phi$ spot) was irradiated for one second. After the laser was turned off, the resistance was reduced by approximately two orders of magnitude, indicating that a conductive patch composed of the low-resistivity charge glass was created near the surface in the matrix of the charge crystal. The emergence of the metastable charge glass can be rationalized by postulating that in response to the activation and deactivation of the SET laser, the local temperature was heated above the charge-ordering temperature $T_{\rm co}$ ($\approx 195$ K) and subsequently rapidly cooled to the environmental (or sample holder) temperature, 100 K [Fig.~2(b), in which inevitable temperature inhomogeneities due to a Gaussian distribution of the laser intensity are ignored for simplicity]. In the RESET process, laser with a relatively weak intensity (80 W/cm$^{2}$) was irradiated for a longer duration (200 s). During the irradiation, the resistance gradually increased with time, which corresponds to a phase conversion from the metastable charge glass to the thermodynamically stable charge crystal at an elevated temperature below $T_{\rm co}$ (see also Supplemental Material Fig.~S1 \cite{Suppl}). When the irradiation ceased, the local temperature reverted to the environmental temperature, and the resistance thus returned to the initial value. The write/erase operation of the low-resistivity charge-glass patch is repeatable [Fig.~2(c)], providing a proof of concept for the rewritable and non-volatile PCM function emerging from correlated electrons.

When considering the performance of the memory function, we found that the ``charge crystallization'' process in $\theta$-RbZn is rather slow, on the order of 10$^{2}$ s. Nevertheless, the slow kinetics appears to be plausible given that the targeted compound $\theta$-RbZn has a low $R_{\rm c}$ ($\approx 8 \times 10^{-2}$ K/s) and given that atomic/molecular liquids with low $R_{\rm c}$ (or, equivalently, high glass-forming ability) generally tend to exhibit low crystallization rates \cite{Ref1}; for instance, the crystallization requires more than 10$^6$ s in the archetypal good glass former SiO$_2$ with $R_{\rm c}$ $\approx 2 \times 10^{-4}$ K/s \cite{Ref1}, whereas the process requires less than 10$^{-8}$ s in the chalcogenide-PCM \cite{Ref9, Ref10}, of which $R_{\rm c}$ is typically $\sim$10$^{9}$-10$^{11}$ K/s \cite{Ref18}. Such a qualitative understanding of the slow kinetics in $\theta$-RbZn led to the working hypothesis that a faster correlated-electron PCM function can be found in a material with a higher $R_{\rm c}$ (or, equivalently, lower ``charge-glass-forming ability'').

\begin{figure}
\includegraphics[width=8.6cm]{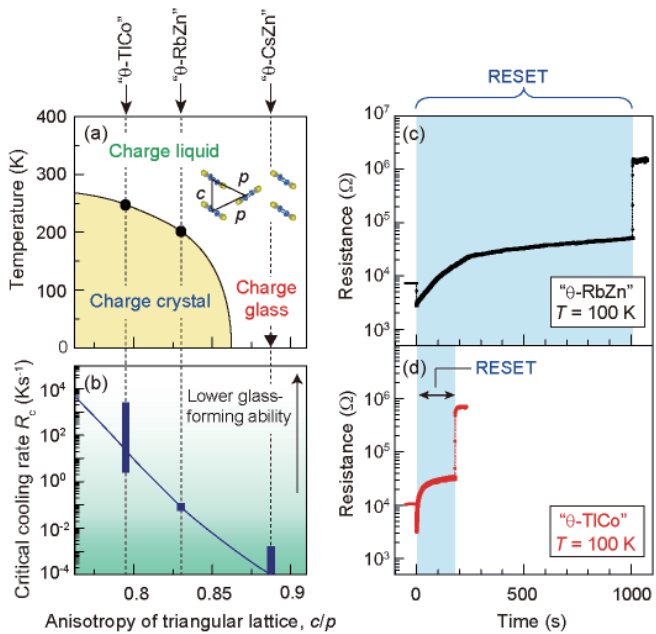}
\caption{\label{Fig3} 
(Color online) (a) Generic phase diagram of $\theta$-(BEDT-TTF)$_2$$X$ plotted with respect to the ratio of the two different triangular lattice parameters, $c/p$. The definitions of the lattice parameters, $c$ and $p$, are illustrated in the inset. 
(b) Critical cooling rate $R_{\rm c}$ as a function of $c/p$. Higher values of $R_{\rm c}$ indicate a lower charge-glass-forming ability. 
(c, d) The time evolutions of the resistance in the charge-crystallization process for $\theta$-RbZn (c) and $\theta$-TlCo (d) under irradiation by the RESET lasers. }

\end{figure}

In pursuing the faster correlated-electron PCM function, we noted that in the isostructural family of $\theta$-RbZn, the charge-glass-forming ability is correlated with the anisotropy of the BEDT-TTF triangular lattice parameters, $c/p$ [Fig.~3(a), inset] \cite{Ref19}; as summarized in Fig.~3(a,b), a $\theta$-RbZn analogue with a more anisotropic triangular lattice, $\theta$-(BEDT-TTF)$_2$TlCo(SCN)$_4$ (denoted by $\theta$-TlCo), exhibits a higher $T_{\rm co}$ ($\approx$ 245 K) and a higher $R_{\rm c}$ ($>$ 2.5 K/s) \cite{Ref19}, whereas the more isotropic one, $\theta$-(BEDT-TTF)$_2$CsZn(SCN)$_4$ (``$\theta$-CsZn''), exhibits no charge crystallization on the laboratory time scale \cite{Ref20, Ref21, Ref22}, presumably owing to the extremely low $R_{\rm c}$ ($<$ 1.7$\times$$10^{-3}$ K/s). Therefore, in terms of $R_{\rm c}$, $\theta$-TlCo with the most anisotropic triangular lattice appears to be a candidate material for the faster correlated-electron PCM function. Although the relatively high $R_{\rm c}$ has thus far precluded kinetic avoidance of the charge crystallization in $\theta$-TlCo experimentally, the present approach using focused laser irradiation produces a local temperature variation at approximately $1 \times 10^{3}$ K/s (see Supplemental Material Fig.~S2 \cite{Suppl}), which was found to surpass the $R_{\rm c}$ of $\theta$-TlCo; the low-resistance state was created in $\theta$-TlCo after irradiation by laser light, analogous to the case in $\theta$-RbZn (see Supplemental Material Fig.~S1 \cite{Suppl}).

Having achieved the SET process in the two isostructural compounds, a comparison of the charge-crystallization time can be now made. Figures 3(c) and 3(d) display the time evolution of the charge-crystal growth in $\theta$-RbZn and $\theta$-TlCo, respectively, under the RESET laser irradiation, whose intensities were each set to the optimal value whereby the maximal ratio of the resistance switching is produced (see Supplemental Material Fig.~S3 \cite{Suppl}). Note that $\theta$-TlCo exhibits faster transformation into the charge crystal; using a fit to a phenomenological relaxation equation (Supplemental Material Fig.~S3 \cite{Suppl}), the characteristic relaxation time $\tau$ can be estimated to be $\approx$11 s for $\theta$-TlCo and $\approx$170 s for $\theta$-RbZn. The one order of magnitude faster $\tau$ in $\theta$-TlCo is consistent with the working hypothesis, demonstrating that finding a material with a lower charge-glass-forming ability can be a guiding principle for developing a faster correlated-electron PCM.

\begin{figure}
\includegraphics[width=8.6cm]{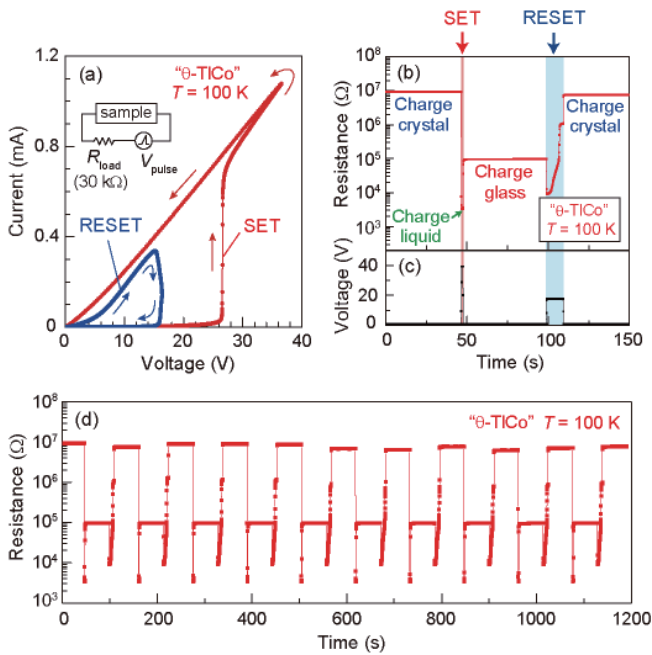}
\caption{\label{Fig4} 
(Color online) (a) The current-voltage characteristics of $\theta$-TlCo connected in series to a load resistor of 30 k$\Omega$, measured under asymmetric triangular voltage pulses. The schematic of the electrical circuit is shown in the inset. 
(b, c) Single-cycle operation of the SET and RESET processes under an application of rectangular voltage pulses. The time profiles of the two-probe resistance and voltage applied to the circuit are shown in (b) and (c), respectively. 
(d) Repetitive operation of the correlated-electron phase-change memory function under the SET and RESET rectangular voltage pulses. In (a)-(d), $\theta$-TlCo was mounted on a sample holder at 100 K.}

\end{figure}

	Finally, we demonstrate that the correlated-electron PCM function is compatible with electrical operation, in which the phase conversion is achieved by harnessing Joule heating. Figure~4(a) shows the current-voltage ($I$-$V$) characteristics of $\theta$-TlCo connected in series to a load resistor ($\approx$ 30 k$\Omega$) that were measured under asymmetric triangular voltage pulses (for details, see Supplemental Material Fig.~S4 \cite{Suppl}). In the SET process, the current through the circuit is initially small but steeply increases when the voltage exceeds some critical value, indicating a switch to a low-resistance state, which persists even after the cessation of the SET pulse; the created low-resistance state can be subsequently initialized by applying an asymmetric triangular voltage pulse with a relatively weak intensity and a long duration (the RESET process). We also found that simple rectangular voltage pulses are sufficient for such resistive switching if an appropriate magnitude and duration are chosen [Fig.~4(b,c)]. The switching displays good repeatability [Fig.~4(d)], indicating that the correlated-electron PCM can be potentially integrated as a component of electric circuits.


Because the concept of correlated-electron PCM function is based on the kinetic avoidance of a first-order charge-crystallization transition, various classes of compounds, including transition-metal oxides as well as organic conductors, can be potential candidates for use in correlated-electron PCM. Given that crystallization rates in atomic/molecular liquids may vary by more than ten orders of magnitude according to their glass-forming ability \cite{Ref1, Ref23}, charge-crystallization rates should be significantly material dependent as well, although little is currently known about this issue. In some materials, the charge-crystallization rate (or the critical cooling rate) may be quite high, and hence, a cooling rate far above 10$^{3}$ K/s may be required to reach quenched electronic states hidden behind long-range ordered states; conversely, such a situation hints at the possibility for a high-speed correlated-electron PCM function once quenched electronic states are stabilized by applying hitherto-untried quenching rates.

	In summary, we have demonstrated PCM functions emerging from a charge-configuration degree of freedom in organic conductors $\theta$-(BEDT-TTF)$_2$$X$ [$X$ = RbZn(SCN)$_4$ and TlCo(SCN)$_4$] . Repeatable switching between a charge-crystalline state and quenched charge glass is achieved via heat pulses supplied by optical or electrical means. $\theta$-TlCo with critical cooling rate $R_{\rm c}$ (2.5 K/s $<R_{\rm c}<$ 2$\times$10$^{3}$ K/s) exhibits switching that is one order of magnitude faster than $\theta$-RbZn with $R_{\rm c}$ $\approx$ 8$\times$10$^{-2}$ K/s, indicating that the material's critical cooling rate can be useful guidelines for pursuing a faster correlated-electron PCM function. Our results establish a clear case whereby practically stable glassy electronic states hidden behind long-range ordered states can be uncovered by adopting hitherto-untried quenching rates and thus underlies a new class of PCM.

	F.K. thank H.~Tanaka and T.~Sato for their fruitful discussions. This work was partially supported by JSPS KAKENHI (Grant No. 25220709).

\end{document}